\documentclass[prb,onecolumn,nopacs,11pt]{revtex4}
\usepackage[english]{babel}
\usepackage{pifont}
\usepackage{amsmath, amsthm, amssymb}
\usepackage{graphicx}
\usepackage{epsfig}
\usepackage{psfrag}
\usepackage{stmaryrd}
\usepackage{color}
\newlength{\upit}\upit=0.1truein
 
\newcommand{\ltappr}{{{\lower4pthbox{$<$} } \atop \widetilde{ \ \ \ }}}

\newcommand \bea {\begin{eqnarray} } 
\newcommand \eea {\end{eqnarray}} 

\newcommand{\dn}{\downarrow}

\newcommand{\beq}{\begin{equation}}
\newcommand{\eeq}{\end{equation}}

\newcommand{\bq}{{\bf{ q}}}
\newcommand{\bk}{{\bf{ k}}}
\newcommand{\bx}{{\bf{ x}}}
\newcommand{\ud}{\mathrm{d}}

\newcommand{\eps}{\epsilon}
\newcommand{\om}{\omega}
\newcommand{\kk}{\mathbf{k}}
\newcommand{\qq}{\mathbf{q}}
\newcommand{\QQ}{\mathbf{Q}}
\newcommand{\RR}{\mathbf{R}}

\newcommand{\boldsigma}{\boldsymbol{\sigma}}
\newcommand{\cks}{c_{\kk\sigma}}

\newcommand{\hks}{\tilde{h}_{\kk\sigma}}

\newcommand{\yrs}{YbRh$_2$Si$_2\,$}
\newcommand{\ybal}{$\beta$-YbAlB$_4\,$}    
\newcommand{\priro}{Pr$_2$Ir$_2$O$_7\,$}
\newcommand{\shs}{Yb$_{2}$Pt$_{2}$Pb}

\def\sgn{\text{sgn}\,}

\def\figx#1#2{\includegraphics[width=#1]{#2}}

\newcommand{\s}{\sigma}

\newcommand{\dg}{^{\dagger}}

\newcommand{\up}{\uparrow}

\newlength{\bxwidth}

\newlength{\fight}
\newcommand{\fg}[3]
{\begin{figure}[here]
\[ 
\figx{\fight}{#1}
\] 
\vspace*{-4mm}
\caption{\label{#2}
\small #3}
\end{figure} }
\def\gtappr{{{\lower4pt\hbox{$>$} } \atop \widetilde{ \ \ \ }}}
\def\ltappr{{{\lower4pt\hbox{$<$} } \atop \widetilde{ \ \ \ }}}

\newcommand{\fgh}[3]
{\begin{figure}[h]\epsfysize=\fight 
\centerline{\epsfbox{#1}}
\caption{{#2}}\label{#3}\end{figure}}
\begin{document} 

\title{Frustration and the Kondo effect in heavy fermion materials}
\author{Piers Coleman$^{1,2}$}
\author{Andriy H. Nevidomskyy$^{1,3}$}
\affiliation{$^{1\,}$Center for Materials Theory, Department of Physics and
Astronomy, Rutgers University, Piscataway, N.J. 08854, USA}
\affiliation{$^{2\,}$Department of Physics, Royal Holloway, University of London, Egham, Surrey TW20 0EX, UK}
\affiliation{$^{3\,}$Department of Physics and Astronomy, Rice University,
Houston, Texas 77005, USA}

\date{\today}

\begin{abstract}
The observation of a separation between the antiferromagnetic phase
boundary and the small-large Fermi surface transition in recent
experiments has led to the proposal that frustration is
an important additional tuning parameter in the Kondo lattice model
of heavy fermion materials.  The introduction of a Kondo (K) and a
frustration (Q) axis into the phase diagram permits us to discuss the
physics of heavy fermion materials in a broader perspective. The
current experimental situation is analysed in the context of this
combined ``QK'' phase diagram.  We discuss various theoretical models
for the frustrated Kondo lattice, using general arguments to
characterize the nature of the $f$-electron localization transition
that occurs between the spin liquid and heavy Fermi liquid
ground-states. We concentrate in particular on the Shastry--Sutherland
Kondo lattice model, for which we establish the qualitative phase
diagram using strong coupling arguments and the large-$N$ expansion.
The paper closes with some brief remarks on promising future
theoretical directions.
\end{abstract}

\maketitle

\vfill \eject

\section{Introduction}

Since the 1980s, the Doniach scenario for heavy fermion behaviour\cite{doniach},
with a single quantum critical point linking the antiferromagnet and the 
heavy fermion metal, has provided the central conceptual framework 
for the understanding of heavy fermion materials. With the growth of
interest in quantum criticality it 
has been tacitly assumed that the single magnetic 
quantum critical point (QCP) predicted by Doniach must uniquely
describe the antiferromagnetic quantum
criticality~\cite{steglichsi,fourfriends} seen in heavy
fermion metals.

Today however, there is a growing 
sense that the Doniach
scenario for heavy fermion behaviour 
may be insufficiently flexible to account for the body of
non-Fermi liquid (NFL) behaviour seen in 
heavy fermion materials. In particular, recent studies of \yrs  under 
pressure\cite{yrs-pressure} and doping with 
Co and Ir~\cite{sven2009} or Ge~\cite{custers10}, point to the existence
of two different transitions as a function of magnetic field -- an
antiferromagnetic (AFM) quantum critical point (QCP) on one hand, and the small-to-large Fermi surface
crossover on the other, manifested by an abrupt jump in the Hall
resistivity. While the position of the antiferromagnetic
transition depends on pressure and doping, the 
small-to-large Fermi surface transition is unaffected 
by changes in pressure and doping~\cite{sven2009}.
In another heavy fermion compound, YbAgGe, several magnetic
transitions are observed as a function of applied magnetic field, and
crucially, the non-Fermi liquid (NFL) state occupies a finite range of
magnetic field, $5<H<10$~T, separating the AFM and the heavy Fermi
liquid phases~\cite{YbAgGe}. 

These experiments  suggest that under certain conditions,
the transition between the fully developed antiferromagnet and the
heavy fermion metal involves two distinct quantum critical points (QCPs), 
or possibly even a line of fixed points forming a \emph{quantum critical
phase}, realized in YbAgGe and doped \yrs as a function of magnetic
field~\cite{YbAgGe,sven2009} and perhaps in \ybal as a function of pressure
\cite{ybal, ybal-scaling}.
By contrast, 
Doniach's classic phase diagram for heavy fermion physics in the Kondo lattice
describes a direct second-order phase transition 
from an AFM phase into a heavy Fermi liquid as the strength of
the Kondo coupling between the conduction and $f$-electrons increases.
While such direct transitions have been observed, for example, in
$CeCu_{6-x}Au_{x}$\cite{hilbert,hilbert2}
and in $CeNi_{2}Ge_{2}$\cite{ceni2ge2}, the more recent results 
indicate that the single-step transition is not universal. 
This line of reasoning leads us to conclude that 
that the phase space
of the Kondo lattice must involve additional variables beyond 
the strength of the Kondo coupling ($K$).

The simplest and most intriguing possibility 
is that the global magnetic phase diagram for heavy fermions 
requires an additional
axis\cite{Bur02.1,Si06,Leb07.1,Voj08.1,Ong09.1,Custers10.1}, 
the  `$Q$' axis, 
which measures the additional quantum zero-point motion of the spins
induced by  magnetic
frustration.
\fg{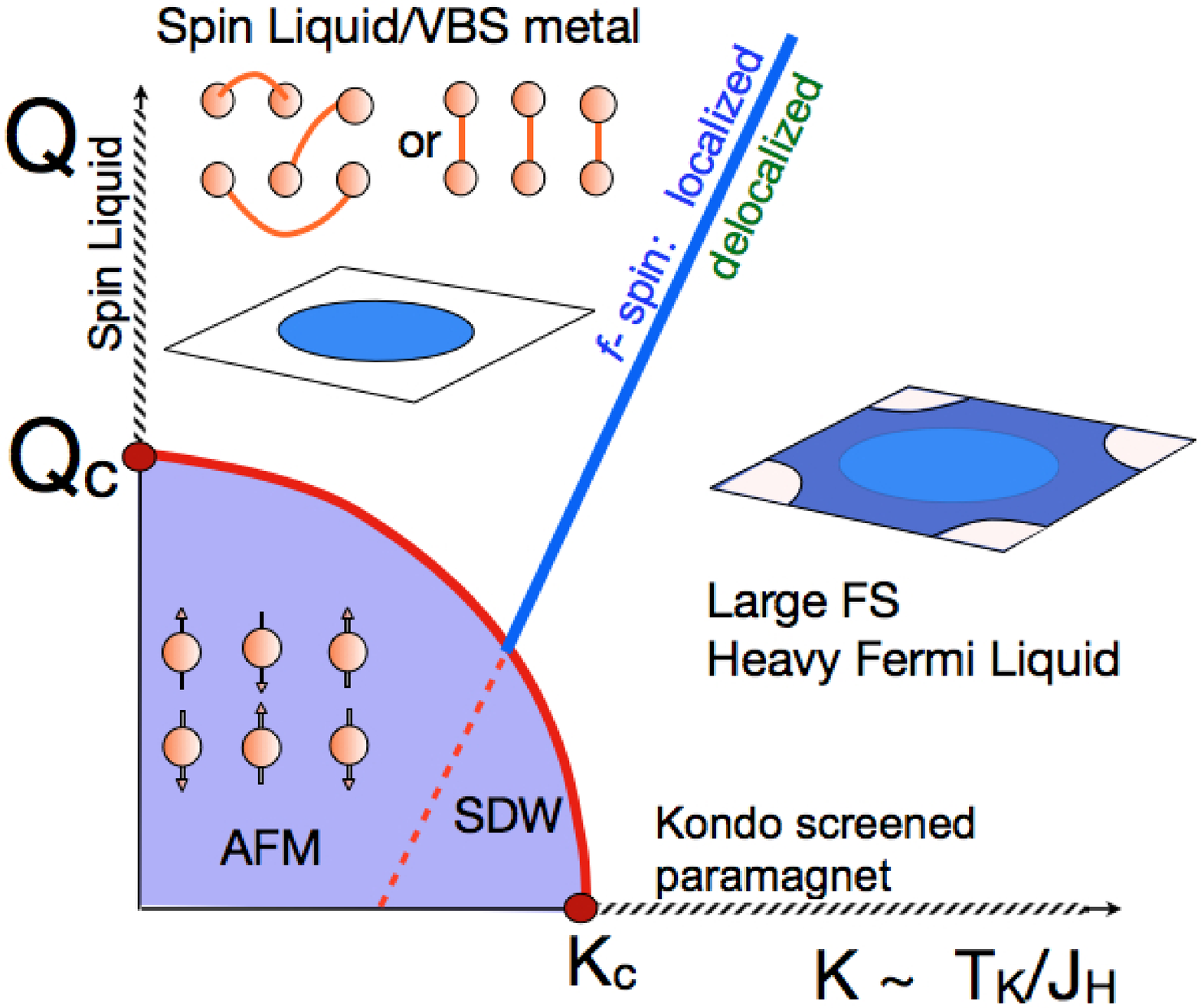}{figbla}{(Colour online) Schematic phase diagram of the
Kondo lattice model in the parameter space of the Doniach axis
$T_K/J^H$ and the quantum frustration axis $Q=1/S$.  A common
antiferromagnetic phase boundary stretches from $K=K_{c}$ on the Kondo
axis to $Q=Q_{c}$ on the frustration axis. At large $Q$, small $K$, a
spin-liquid metal with localized $f$-electrons and a small Fermi surface
forms, whereas at large $K$, a heavy Fermi liquid with a large Fermi
surface and delocalized $f$-electrons develops. Since the volume of the
Fermi surface is conserved, it is not
possible, on a variety of lattices, to evolve smoothly from a small to a large
Fermi surface, so that the spin liquid metal and the heavy Fermi liquid must be
separated by a zero-temperature quantum phase transition. }

In this paper we explore and review this idea.
The two-axis diagram describing the joint effects of the
Kondo screening ($K$) and quantum zero-point motion ($Q$) we call the ``$QK$''
diagram, shown in its simplest form in Fig \ref{figbla}.
To illustrate the idea, first 
consider draining away the mobile electrons in a heavy fermion material to
reveal the underlying magnetic lattice of $f$-electrons, 
($K=0$), each forming local moments coupled together via 
short-range antiferromagnetic (AFM) Heisenberg interactions of
characteristic strength $J^{H}$.  
We then
reintroduce the mobile electrons, and 
consider the effect of tuning up their coupling $K$
to the underlying magnetic lattice, 
as illustrated in Fig. \ref{fig2l}

The new
element, is that this lattice of local moments is considered to be 
magnetically frustrated.
The appearance of frustration in real heavy  fermion systems may take
various guises -- in certain cases, it can appear as direct geometric
frustration, as in the pyrochlore heavy fermion material \priro
\cite{nakatsuji-PrIrO} and the Shastry--Sutherland lattice compound, \shs
\cite{kim08}.
Frustration can also take other forms, derived from competing
interactions of various kinds. For example, in the ``heavy fermion''
physics of bilayer He-3, frustration may derive from ring-exchange
effects in the lower, almost localized layer of helium
atoms\cite{saunders2007}. 
\fg{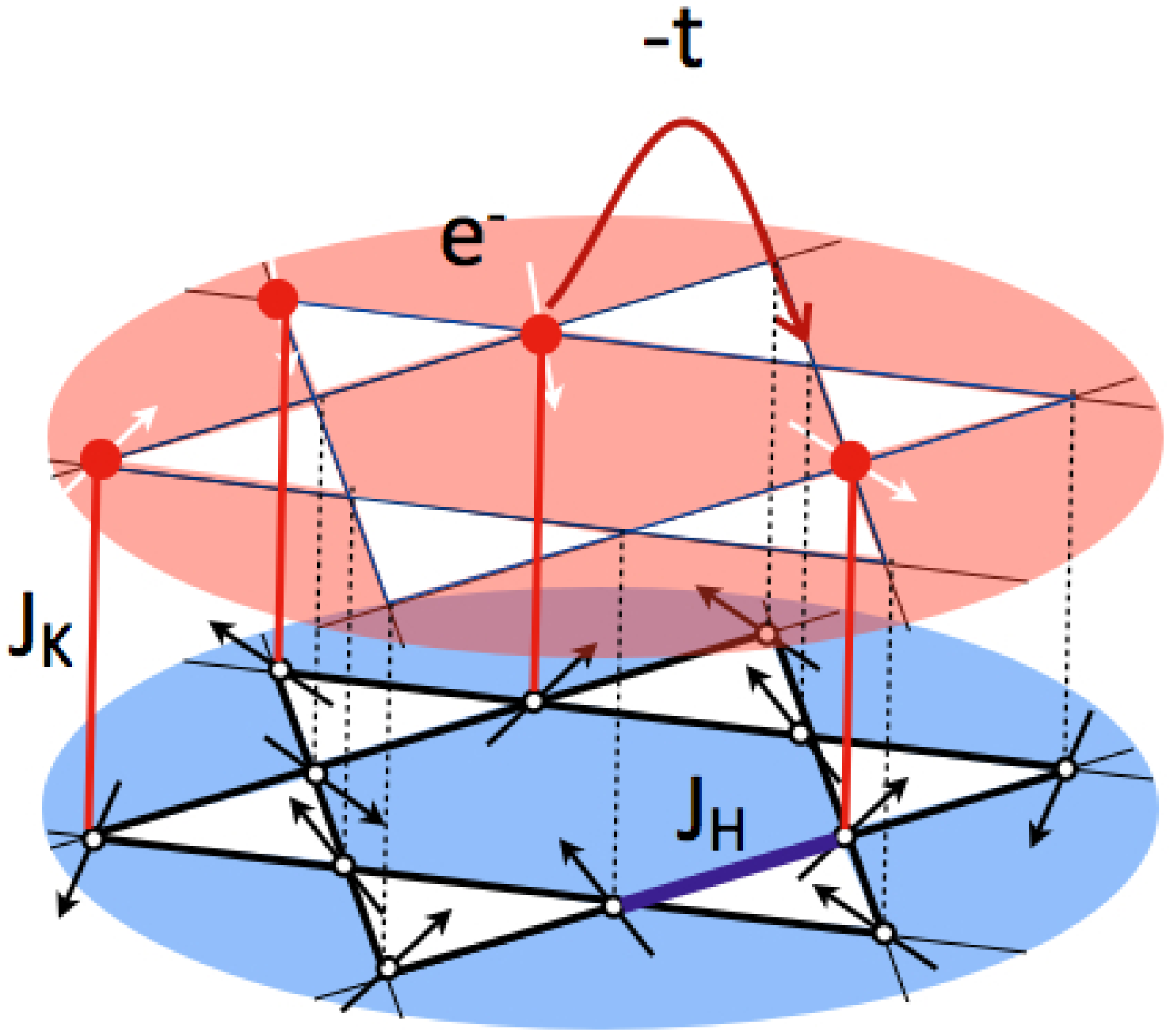}{fig2l}{(Colour online) Illustrating the Kondo
coupling ($J_{K}$) between  mobile
electrons and an underlying frustrated spin system in a hypothetical
Kondo-Kagom\'e lattice.}

The staggered order-parameter of a quantum antiferromagnet
$\vec{S}_{\bf{ Q}}$ does not commute with the Hamiltonian, giving 
rise to zero-point fluctuations in the magnetization. 
In stable antiferromagnets these fluctuations act to reduce the size of
the local moment below its classical value\cite{anderson52}, 
but by increasing the frustration, the zero
point motion can be driven so high that at some critical value
$Q=Q_{c}$, the 
antiferromagnetic order melts, forming a spin liquid or a valence bond
solid of spin dimers\cite{fazekas74,SS}. Depending on the
lattice, this tuning can be done in various ways as illustrated in
Fig. \ref{fig1l}.
For linguistic 
convenience, we shall loosely refer to this region as the ``spin-liquid region''
of the $QK$ diagram.

Antiferromagnetic order can also be destroyed by the screening
effects of the Kondo physics.  According to the Doniach scenario, if
reintroduce the electrons to the Kondo lattice, once the Kondo
temperature becomes comparable with the RKKY coupling between the
moments, $K\equiv T_{K}/J^{H}=K_{c}\sim 1$, the screening of the local
moments becomes complete, and a quantum phase transition into a heavy
Fermi liquid takes place. In the $QK$ diagram, this is the limiting
behaviour along the $x$-axis ($Q=0$), where frustration is absent.

The key idea of the $QK$ diagram is to unify the Kondo and frustration
effects as shown in Fig.~\ref{figbla}, in which the two
limiting quantum critical points at $Q=Q_{c}$ and $K=K_{c}$ are linked
by a single antiferromagnetic phase boundary.  Outside this phase
boundary, at large $K>K_{c}$ but small $Q$, the system is a heavy Fermi
liquid, in which the local magnetic moments are fully screened,
donating their spin degrees of freedom to the Fermi sea to form a
large Fermi surface of heavy electrons.  By contrast, at small $K$ and larger
than critical frustration $Q>Q_{c}$, the localized spins form a metallic
spin liquid (or valence bond
solid), and the conduction electrons are decoupled from the spin
fluid, forming a metal with a small Fermi surface.  Since the size of
the Fermi surface is an adiabatic invariant, this leads us to conclude
that there is no continuous way to move from the spin liquid metal to
the heavy Fermi liquid.  This leads to the tentative conclusion that
at the very least, there must be one or more zero-temperature phase
transitions separating the heavy Fermi liquid from the spin liquid
metal.

In this paper we examine this reasoning in greater
detail in relationship both to concrete theoretical models  and
real materials.  One of the unexpected surprises, is that 
the phase boundary between the spin liquid and the heavy Fermi liquid is only
guaranteed to exist
in lattices where the unit cell contains an odd number of local moments.
At the end of the article we discuss the relationship of the $QK$
diagram to current experiments and ongoing efforts to theoretically
understand the nature of heavy fermion quantum criticality. 

\section{The QK diagram.}
\subsection{Kondo Screening vs Zero-point motion}\label{screening}

\fight=\textwidth
\fg{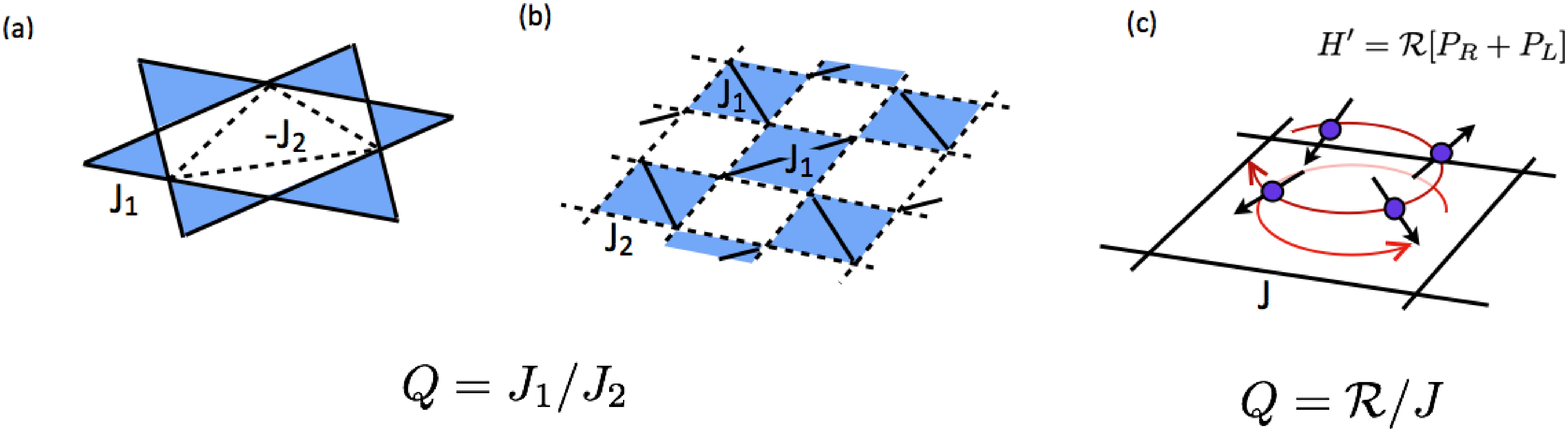}{fig1l}{(Colour online)
Frustrated Heisenberg models with a tunable degree of frustration $Q$. (a)
The  geometrically frustrated $S=1/2$ kagom\'e lattice, with next
nearest neighbour ferromagnetic couplings to stabilize
antiferromagnetism, and 
(b) the Shastry--Sutherland model, with alternating diagonal antiferromagnetic
interactions $J_1$ on the a checker-board array and a nearest neighbour
interaction of strength $J_{2}$.
For (a) and (b), when  $Q=J_{1}/J_{2}$
is large, the antiferromagnetism is destroyed, leading to a 
spin liquid in the kagom\'e lattice, and a valence bond solid
across the diagonal bonds in the Shastry--Sutherland lattice, respectively. (c) 
The 2D Heisenberg antiferromagnet with a spin ring exchange term\cite{sandvik02}
Eq.~(\ref{ring}). When 
$Q={\cal R}/J$ is increased, the two-sublattice
magnetization is destroyed.
}
\fight=0.8\textwidth

One of the main ideas of $QK$ phase diagram, is the existence 
of two distinct mechanisms for the destruction of antiferromagnetism in a Kondo
lattice. We begin with a discussion of this idea. 
If we drain the electrons out of a Kondo lattice, we are left with 
a system of localized moments that interact with each other
via an arbitrary Heisenberg interaction $J^H$, plus perhaps the ring-exchange
terms:
\begin{equation}
\hat{H}_f = \sum J^H_{ij}\, \hat{\mathbf{S}}_i \cdot
\hat{\mathbf{S}}_j +\hat{H}_{\square},
\label{Hspin}
\end{equation}
In some cases it may be necessary to consider an
additional spin ring exchange effect:
\beq\label{ring}
\hat{H}_{\square} = {\cal R}\sum_{\hbox{plaquettes}}[\hat{P}_{R}+\hat{P}_{L}]
\eeq
where $\cal R$ is the amplitude for ring exchange of spins around a
plaquette, and $\hat{P}_{R,L}$ are the operators that exchange spins in a
right, or left-handed sense around the plaquette.

In heavy fermion systems the localized moments develop in
partially filled 4$f$ or 5$f$ shells. 
Unlike the classic examples of magnetic frustration,
heavy fermion materials are metals, and  hybridization
with conduction electrons plays crucial role.  In well-localized heavy
electron systems, the hybridization between $f$- and conduction
electrons gives rise to a 
Kondo (antiferromagnetic) on-site
interaction $J_K$ between the spins of the conduction and
$f$-electrons.   Once we re-introduce  conduction electrons, the 
combined Hamiltonian becomes 
\begin{equation}
\hat{H} =
\hat{H}_f + \hat{H}_K,
\end{equation}
where
\begin{equation}
\hat{H}_K = \sum_{\kk\s} \eps_k \cks\dg \cks + 2J_K 
\sum_{i}
\hat {\bf s}_{c} (i)
\cdot \hat{\mathbf{S}}_i
\label{H_cf}.
\end{equation}
describes the conduction sea and its coupling to the localized
moments. 
Here $\hat {\bf s}_{c} (i)\equiv \frac{1}{2} (c_{i\alpha}\dg 
\boldsymbol{\sigma}_{\alpha\beta}\, c_{i\beta})
$ describes the spin density 
of conduction electrons at site $\RR_i$ through electron
creation operators $c^\dag_{i\s} = \sum_\kk \cks\dg
e^{i\kk\cdot\RR_i}$ and $\boldsymbol{\sigma}_{\alpha\beta}$ are the Pauli
matrices.

If a magnetically ordered phase exists, the size of the ordered
(condensed) moment can be expressed through the identity:
\begin{equation}M_0^2 =\left\langle  \left(\hat{\mathbf{S}}_f +
\hat{\mathbf{s}}_c\right)^2 \right\rangle  -
3\int\frac{\ud^d q}{(2\pi)^d} \int\limits_0^\infty \frac{\ud
\omega}{\pi} \left(\frac{1}{2} + \mathrm{n}(\om)\right)
\chi_\text{tot}''(\qq, \om),\label{M0} \end{equation} where $\mathrm{n}(\om)$ is
the Bose distribution function, and 
$\chi''_\text{tot} $ 
is the dynamical susceptibility of
the magnetization
\begin{equation}\label{chi}
\chi''_{tot} (q)\delta_{ab}= \frac{1}{2}\int \ud^{4 }x \langle  [ \hat M_{a}
(x),\hat M_{b}(0)]\rangle e^{-iqx}, \qquad  (q\equiv (\bq ,\omega),\ qx \equiv \bq  \cdot
\bx-\omega t).
\end{equation}
Here $\hat {\bf M }(x)= \hat{\bf s}_{c}+ \hat {\bf S}_{f}$ is the
total magnetization. The dynamical susceptibilty can be expanded in
terms of both localized and conduction electron
contributions $\chi''_\text{tot} = \chi''_{cc} + \chi''_{ff} +
2\chi''_{fc}  $. 

Equation~(\ref{M0}) offers insight into two alternative ways in which
the size of the ordered moment may be reduced:
\begin{itemize}
\item  {\bf Spin zero-point motion}, encoded in the
second term on the r.h.s. of Eq.~(\ref{M0}) which reduces the
magnitude of the mean-field ordered moment by transfering spin spectral
weight from the ordered moment into  the finite frequency part of the
spin fluctuations. In a pure insulating  antiferromagnet, the $f$-electron
ordered moment will be, generally, smaller than the classical value
$S$, as follows from Eq.~(\ref{M0}):
\begin{equation}
M_f^2 = S(S+1) - 3\int\frac{\ud^d q}{(2\pi)^d} \int\limits_0^\infty \frac{\ud \omega}{\pi}  \left(\frac{1}{2} + \mathrm{n}(\om)\right) \chi_{ff}''(\qq, \om),
\label{Mf}
\end{equation} 
{\emph{Geometric frustration}} enhances zero-point motion of
spins,  transfering spectral weight from the spin-condensate into the
fluctuations described by $\chi_{ff}'' (\bq ,\omega)$. This 
reduces the size of the ordered moment, and beyond a critical value of
frustration $Q_{c}$, 
no long-range spin order can survive. Examples include
the kagom\'e
lattice or square lattice with particular ratio of next-nearest
neighbour interaction $J_2/J_1 = 1/2$. In all these cases, the
zero-point motion is con\-cen\-trated in a region of momentum space 
surrounding the classical ordering wave-vector $\QQ$. 

\item {\bf \emph{Kondo screening}}. This effect can be simply
understood as reduction in the first term on the r.h.s. of
Eq.~(\ref{M0}), due to the antiferromagnetic interaction between the
localized electron moment $\mathbf{S}_f$ and the conduction electron spin
$\mathbf{s}_c$.  Note that the integral in the last term in Eq.~(\ref{M0})
will also decrease, because of the negative contribution of
hybridization term $\chi_{cf}$ into the total susceptibility
(\ref{chi}).  It is the mixed component of the susceptibility that, at
temperatures below the Kondo temperature $T_K$, encodes the Kondo
screening, leading to the reduction of the localized moment $S_f\to
S-1/2$ in the ground state. The important difference from the
zero-point motion discussed above, is that $\chi_{cf}''(\qq,\omega)$
is a local quantity and unlike the effects of spin zero-point motion, 
the effects of Kondo screening are diffusely distributed in momentum space. 
\end{itemize}

In fact, both zero-point fluctuation and spin-screening effects 
must be present in a heavy fermion compound, providing the 
basis for two independent axes in the
generalized $QK$ heavy fermion phase diagram, shown in Fig. \ref{figbla}.
The $x$-axis  describes the Kondo screening, with a tuning parameter
given by the ratio $K = T_{K}/J^{H}$ of the Kondo temperature $T_{K}$ 
to the characteristic antiferromagnetic coupling strength $J^{H}$.
The $y$-axis describes the effect of spin zero-point motion, 
tuned by the frustration parameter $Q$. We now discuss this diagram in
detail, with reference to various concrete model examples. 

\subsection{General considerations}\label{General}

We now turn to a more detailed discussion of the ``QK'' diagram,
combining Kondo screening ($K$) and frustration ($Q$) as the $x$ and
$y$-axes.  Let us first discuss the axes of this diagram. 
In his semi-qualitative phase diagram, 
Doniach~\cite{doniach} argued that the transition
from magnetically ordered phase into a paramagnetic heavy Fermi liquid
occurs as a function of increasing Kondo coupling $J_K$ at a point
where the Kondo temperature becomes comparable with the strength of
the $RKKY $ interaction $K_{c}= T_{K}/J^{H}\sim 1$. In a realistic model
the RKKY scale responsible for long range magnetic order scales as $J^H\sim
J_K^2$, while the Kondo scale is exponential in $J_K$ and eventually wins as
the latter increases, $T_K\sim D \exp(-D/2J_K)$ (where $D$ is the conduction
electron bandwidth). An equivalent way to think of Doniach's phase
diagram is to assume that the RKKY strength and the Kondo temperature
are independent quantities, while their ratio $K= T_K/J^H$ is the
relevant parameter, as plotted on the horizontal axis in
Fig.~\ref{figbla}.
 
The $y$-axis is the frustration axis. 
In real systems, the strength of zero-point motion $Q$ can be
associated with frustration, associating it for example, with the ratio of
competing  bond strengths. From a theoretical stand-point, 
the enhancement of zero-point spin fluctuations can also be
accomplished by considering a tuning of the  spin $S$
of the local moments.  In practice this is done in a large-$S$
or a large-$N$ expansion using a Schwinger  boson description of the
local moments. In these more theoretical approaches, 
we can loosely identify $Q\sim N/ (2S)$ where $N$ is the number of components
of a SU($N$) or Sp($N$) spin. There is a small caveat
associated with these approaches, for once the size of the Kondo spin
becomes greater than $S=1/2$, we require $2S$ screening channels to
fully screen the local moment when the Kondo coupling is turned on,
requiring that a large-$N$ or large-$S$ expansion be done with a Kondo
lattice in which there are then $k=2S$ conduction bands, each
individually screening the local moments\cite{rech06,lebanon07}. 

In the absence of coupling to conduction electrons (i.e in insulating
magnets), the effect of quantum frustration is relatively well
understood. For example, in spin-wave theory, the magnetic order
disappears above some critical value $Q_c=1/S_c$ (on a square lattice
$S_c\sim 0.2$)\cite{anderson52}.  
At higher values of $Q$, a dimerized valence bond solid (VBS)
phase,  or alternatively  \emph{spin liquid} -- a state with no
broken symmetry, are expected to develop.  

Once we combine the Kondo ($K$) and frustration ($Q$) effects into a
single diagram, antiferromagnetism occupies the small $K$, small $Q$
corner of the phase diagram. Since there is only one antiferromagnetic
phase, there must be a single phase boundary that connects the quantum
phase transitions at $K=K_{c}$ on the x-axis and $Q=Q_{c}$ on the
y-axis.  

The paramagnetic phases that exist outside the AFM phase boundary
differ qualitatively at small and large $K$.  At large $K$, the
quenching of the localized moments in the Kondo lattice liberates
spin into the conduction sea,  and we expect a Fermi surface volume $V_{FS}$
determined by
\begin{equation}
\frac{V_{FS}}{(2\pi)^{D}} = n_{e } + \frac{n_{s}}{2}
\end{equation}
where $n_{e}$ is the density of electrons per unit cell per spin index while
$n_{s}$ is the number of spins per unit cell. This expresses the
fact that each spin liberates one heavy electron degree of freedom, or
half an electron per spin index. 
Now at small finite $K$, 
the Kondo effect will not take place so that 
$V_{FS}/ (2 \pi)^{D} = n_{e}, $
implying that
\begin{equation}
\frac{\Delta V_{FS}}{(2\pi)^{D}} =  \frac{n_{s}}{2}
\end{equation}
between small and large $K$ phases.
Now, the volume of the Fermi surface is only defined modulo
$(2\pi)^{D}$, so an increase in the electron count per unit
by $n_{S}/2$ will lead to a change in the Fermi surface volume only if
$n_{S}$ is an odd number. In this case, provided no symmetries are broken,
the large and small $K$ Fermi surfaces cannot be connected
continuously, and must be separated by a (zero temperature) quantum
phase transition. Of the five well-known lattices, the square,
triangular and kagom\'e lattices have an odd number of spins per unit
cell and are expected to exhibit the small-to-large Fermi surface transition.
By contrast, the Shastry--Sutherland and pyrochlore lattices have
four spins per unit cell and thus cannot be guaranteed to exhibit a
phase transition between  small and large $K$ paramagnetic phases:
\[
\begin{tabular}{l|c|c}
\hbox{Lattice} & \hbox{ Spins/u.$\,$cell } & \hbox{  spin-liquid-FL QCP}\\
\hline
\hbox{Square/Triangular } & 1&  \checkmark \cr
\hbox{Kagome} &3 &  \checkmark\cr
\hbox{Shastry-Sutherland }& 4 & \hbox{x}\cr
\hbox{Pyrocholore} & 4& \hbox{x}
\end{tabular}
\]
Most studies of the frustrated  square lattice Heisenberg model
suggest that rather than a spin liquid, a valence bond solid forms
in the presence of strong frustration, breaking the lattice symmetry. 
In this case, the phase boundary to the heavy Fermi liquid will extend
to finite temperature.  The important point however, is that 
under quite general conditions, provided there is an odd number of
spins per unit cell, in the absence of antiferromagnetic order, 
there must exist some kind of quantum phase
transition between the small- and large- Fermi surface state. 

This is of course, a {\sl minimal requirement}, which does not
preclude more complex transitions between the small and large Fermi
surface states. As we shall see, this may even be possible in those
cases with an even number of spins per unit cell.  To address these
questions requires a more microscopic approach to the Hamiltonian.
The important point is that from the $QK$ diagram, we are able to deduce
that in a Kondo lattice with
an odd number of spins per unit cell, 
the ``delocalization'' line for the $f$-electrons 
must separate from the antiferromagnetic phase
transition at large enough frustration parameter $Q$.  This has two qualitative
consequences:

\begin{itemize}
\item 
{\bf Kondo stabilization of the ``spin liquid''}.  The qualitative
form of the AFM phase boundary in the $QK$ diagram raises the
fascinating possibility that once $K>0$, Kondo screening reduces the
size of the local moment, reducing 
the critical value $Q_{c}$ for the formation of a spin liquid. 
This ``Kondo stabilization'' of the spin liquid state was first
suggested by Coleman
and Andrei in their 1987 adaptation of the RVB pairing idea to heavy fermions.\cite{coleman-andrei}.

\item {\bf Local moment-spin-density wave transition.} 
It is very natural to assume that when the $f$-electron delocalization line
meets the magnetic phase boundary, it continues on inside the magnetic phase.
The appeal of this conjecture\cite{Si06}, is that it
permits a separation of the AFM phase into  ``localized'' and a
``spin-density wave'' (SDW) regions, allowing for
the possibility of heavy fermion materials in which the quantum
critical point is described by an itinerant,  Hertz--Millis
scenario\cite{hertz,millis}. This 
boundary may take the form of a  Lifschitz transition in the
underlying Fermi surfaces.

\end{itemize}

Beyond these general considerations, very little is known about the
detailed form of the global phase diagram, and a number of variant
forms have been proposed (Fig. \ref{fig.scen}), largely motivated
by experiment.  One possibility, motivated largely by experiment, is
that the antiferromagnetic phase boundary and the spin delocalization
line merge over finite region of the phase diagram\cite{Si2010} as illustrated in
Fig.~\ref{fig.scen}(a), allowing for the
possibility of a region of the phase boundary governed by a
``local quantum critical point'' where the antiferromagnetism
and the localization of the $f$-electrons occurs simultaneously.
A second possibility, also motivated by experiment, 
is that the transition between small and large
Fermi surface metal takes place via an intermediate ``strange metal''
phase depicted in Fig.~\ref{fig.scen}(b).
We shall later return to discuss these scenarios in the context of
experimental observations.
\fight=0.45\textwidth
\fgh{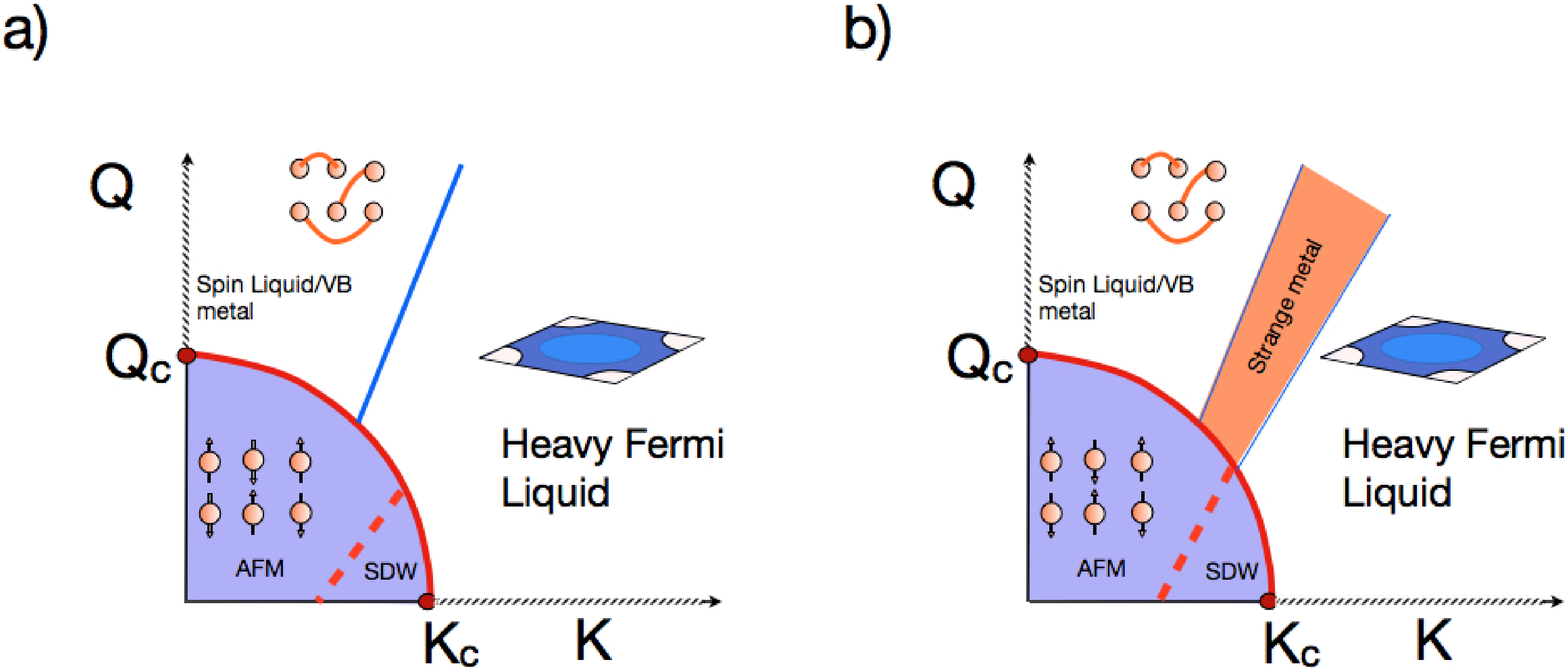}
{(Colour online) Two alternative scenarios for the global $QK$
phase diagram: (a) in which the $f$-electron delocalization line merges
with the antiferromagnetic boundary over a finite region\cite{Si2010}, (b) in which
the transition between small and large Fermi surface takes place via
an intermediate ``strange metal'' phase. }{fig.scen}
\fight=0.8\textwidth

\subsection{The Shastry--Sutherland Kondo lattice}\label{sec.SSK}


\begin{figure}[!tb]
\begin{center}
{\includegraphics[width=0.48\textwidth]{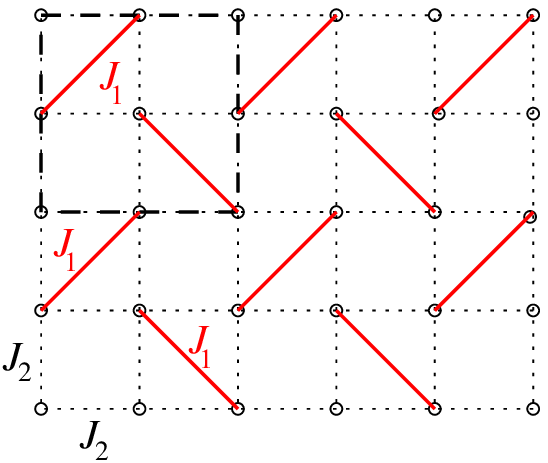}}
\end{center}
\caption{(Colour online) Shastry--Sutherland lattice with each vertex occupied
by a localized
electron spin. The next nearest neighbours interact via  Heisenberg
interaction $J_1$ along alternate diagonals, favouring the dimerized ground
state,
while the nearest neighbour interaction $J_2$ favours the N\'eel
antiferromagnet. The 2x2 elementary unit cell is shown with dashed line,
corresponding to the reduced Brillouin zone in $\kk$-space.
 } 
\label{Fig.Shastry}
\end{figure}

Here  we discuss a particularly simple example of a
frustrated Kondo model: the ``Shastry--Sutherland'' Kondo lattice. 
This lattice is obtained by Kondo-coupling conduction electrons 
on a square lattice to a frustrated Shastry--Sutherland (SS) spin
model~\cite{SS}.  This latter is a checker-board lattice
of
Heisenberg $S=1/2$ moments, in which the coloured squares of the
checker board contain alternating diagonal bonds of strength $J_{1}$, as shown
in Fig.~\ref{Fig.Shastry}. Each
vertex of the SS lattice is thus connected to one partner on the diagonal by
$J_1$ (thick solid lines) and to 4 nearest neighbours by $J_2$ (thin dotted
lines).
The frustration parameter is the ratio $J_{1}/J_{2}$ of the diagonal
to the nearest-neighbour antiferromagnetic couplings $J_{2}$.
Recently, an experimental realization has been found in a
quasi-two-dimensional heavy fermion system~\cite{kim08} Yb$_2$Pt$_2$Pb, with Yb$^{3+}$ ions
forming the Shastry--Sutherland lattice, and in its
Ce analogs~\cite{kim2010}. 

From the theoretical stand-point, the Shastry--Sutherland lattice
has a clear advantage
that the spin-ground-state at large frustration is a
valence-bond solid with spin-singlets arranged on the diagonal bonds
and a well-defined wave-function. Unlike other ``spin liquid''
ground-states, this dimer state was proven to be the \emph{exact}
ground state~\cite{SS} of the SS model (provided $J_1\gtrsim 2 J_2$) and in
particular, is well captured by large-$N$ expansion. It is to our
knowledge, the only frustrated Kondo lattice where the large-$N$ expansion can
reliably examine the effect of Kondo coupling. This is therefore an ideal
starting point to examine the combined effects of the Kondo screening and strong
frustration.

  The Hamiltonian of
the
Shastry--Sutherland Kondo lattice model is 
\begin{equation}\label{SSK1}
\hat{H}_{SSK} =\hat{H}_{K}+ \hat{H}_{SS}\,,
\end{equation}
where 
\begin{eqnarray}\label{SSK2}
\hat{H}_K &=& \sum_{\kk\s} \eps_k \cks\dg \cks + 2J_K 
\sum_{i}
\hat {\bf s}_{c} (i)\cdot \hat{\mathbf{S}}_i\cr
\hat{H}_{SS}&=&
J_{1}\sum_{\boxslash,\boxbslash}\mathbf{S}_k\cdot\mathbf{S}_l+
J_{2}\sum_{\langle i,j \rangle}\mathbf{S}_i\cdot\mathbf{S}_j 
\end{eqnarray}
describe the Kondo and magnetic parts of the Hamiltonian. Here
$\boxslash, \boxbslash$ refers to the sum over plaquettes with alternating
diagonal bonds.  The dispersion of the conduction sea is determined by
a tight binding model on a regular square lattice, with hopping of strength $t$
between nearest neighbour sites, so that $\epsilon_{\bk }= -2t (\cos k_{x}+ \cos
k_{y})-\mu$. In the SS lattice, the ratio of the
diagonal to the nearest neighbour interaction, $Q=J_{1}/J_{2}$, plays the role
of the
frustration parameter.

In the absence of Kondo screening, $J_K=0$,
Shastry and Sutherland\cite{SS} showed 
that the frustrated lattice has a dimerized ground state at large
$Q$, 
\begin{equation} 
|\Psi_\text{SS}\rangle = \prod_{\boxslash,\boxbslash} |d\rangle_{kl}, 
\end{equation}
where $|d\rangle_{kl} = (|\!\!\up_k\dn_l\rangle
-|\!\!\dn_k\up_l\rangle)/\sqrt{2}$ defines a dimer (singlet) on the
diagonal bond. In the opposite limit, $Q\ll 1$, the
ground-state is a N\'eel AFM. The transition between the
two phases is still controversial.  High-temperature series
expansion~\cite{SS.series-expansion}, exact
diagonalization~\cite{SS.ED} and variational
studies~\cite{SS.variational} point to a direct transition between the
dimer phase and the N\'eel ground state at $Q_{c}=(J_{1}/J_{2})_c =
1.43\pm0.02$. On the other hand, a number of
studies~\cite{SS.albrecht-mila,SS.koga,SS.chung-sachdev,SS.hajj}
suggest that an intermediate phase exists between
$1.1\lesssim Q\lesssim 1.65$ with different proposals as to its
nature, including a helical AFM, a plaquette singlet phase or a
columnar phase.

The SS lattice has four spins per unit cell and following the counting argument 
of the last section, \emph{a priori}
we might expect a continuous evolution from the 
the small Fermi surface metal to large Fermi surface heavy electron state.

However, there is a limit in which a phase transition does occur:
the half-filled lattice. With one conduction
electron per site, the $K=0$ state is a metal, whereas the $K=\infty $
state is a Kondo insulator, implying 
a metal-insulator transition at half 
filling  which separates the small and large $K$ limits. Since there is no
change in symmetry, this metal-insulator transition will be a 
{\sl zero temperature}  quantum critical point, with gapless charge excitations.

When we dope away from half filling, we expect
the quantum critical point 
to influence the physics at finite doping, giving rise to instabilities.
Indeed, as we now argue, based on both strong coupling
arguments and the large $N$ expansion, at low enough temperatures
the large and small Fermi surface phases are likely to be unstable to
a $d_{xy}$-wave superconducting instability induced by the magnetic coupling
along the diagonal bonds.  To this end, we will limit our discussion to the
limiting case of infinite frustration where $J_{2}=0$ ($Q=\infty $). 

Consider first the strong-coupling limit of the Shastry--Sutherland Kondo
lattice, in which both the Heisenberg and Kondo couplings are much
larger than the hopping, ${J_{1},\ J_{K} \gg t}$.  We can consider two
extreme limits:

(i) Weak Kondo coupling,  $J_{1}\gg J_{K}\gg t$.   It is useful to
visualize the situation using an RVB notation to describe the
configuration of the singlet bonds between the electrons and the
localized moments. Single electrons will have their energy lowered by
an amount of order $-J_{K}^{2}/J_{1}$ by inducing a virtual resonances
of the dimer bond in which one end of the bond re-attaches to the conduction
electron, forming a Kondo singlet.
This effect can only occur with a singly-occupied conduction electron site, and
will be absent with two electrons above the same site, thereby
generating an effective repulsive on-site $U\sim J_{K}^{2}/J_{1}$ between
electrons. By contrast, when two electrons `hover' above  
a diagonal valence bond, that valence bond can resonate
into the conduction sea.  In the second order perturbation
theory, this leads to an induced antiferromagnetic interaction between
conduction electrons of strength $g\sim J_{K}^{2}/J_{1}$ across
the alternate diagonals of the electron lattice, so that the effective
model describing the electrons at small $K$ is given by
\begin{equation}
H_{eff} = \sum_{\kk\s} \eps_k \cks\dg \cks +
U\sum_jn_{j\uparrow}n_{j\downarrow}+ g
\sum_{\boxslash,\boxbslash} \boldsigma_k\cdot\boldsigma_l
\end{equation}
where $g, U\sim J_{K}^{2}/J_{1}$. 
The weak antiferromagnetic interaction between the electrons across
the diagonals of the alternate plaquettes will couple to the divergent
Cooper pair instability in the $d_{xy}$ channel, giving rise to a
Cooper instability into a weakly paired state with $d_{xy}\sim \sin k_{x}\cdot
\sin k_{y}$
symmetry (note that repulsive $U$ will suppress the competing instability
into an extended s-wave state).  In general all other competing instabilities
will not have a divergent susceptibility, and will be absent for
small enough $J_{K}/J_{1}$.

\fg{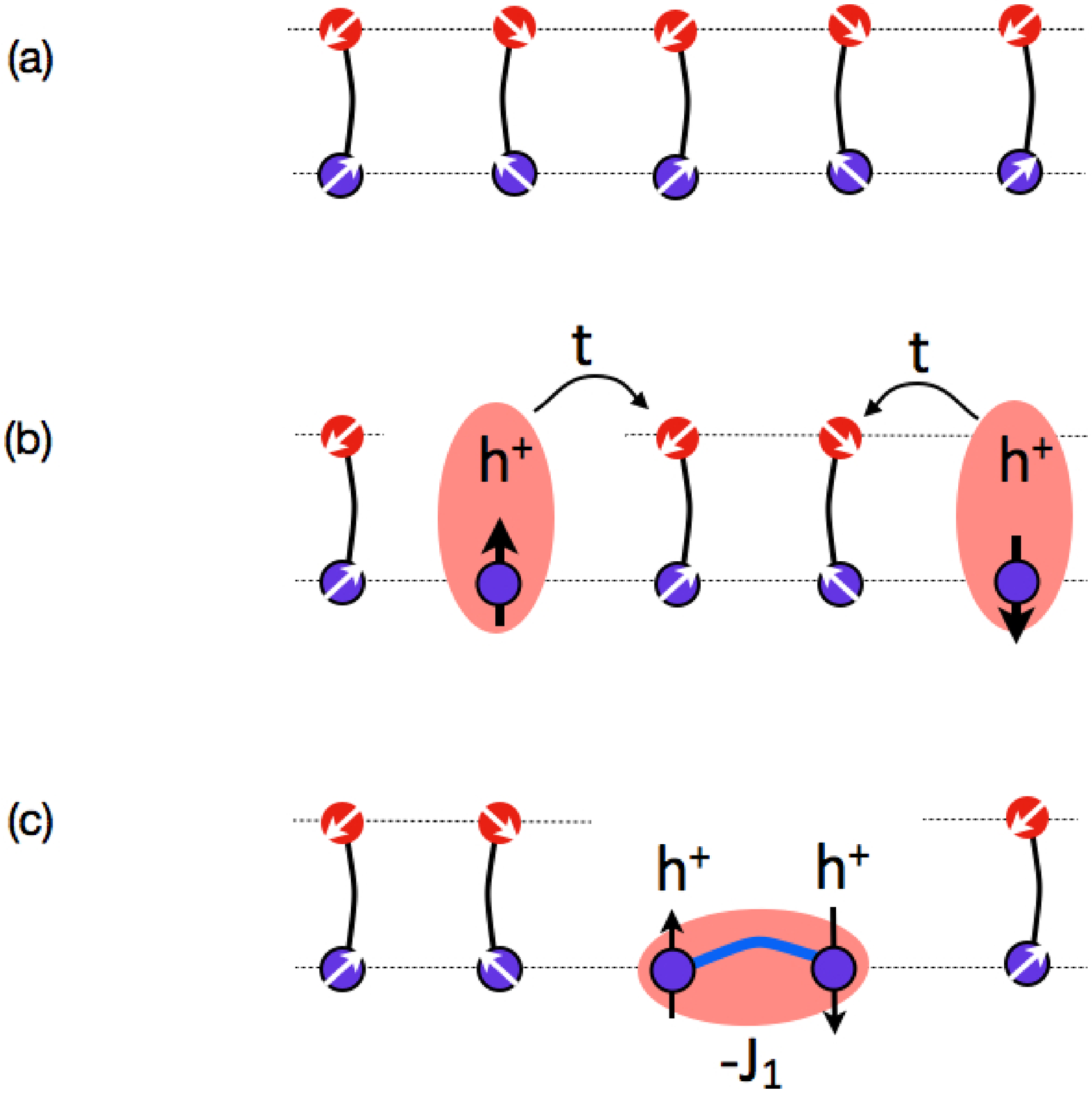}{fig.strong}
{(Colour online) (a) In the strong coupling limit $J_K \gg J_1 \gg  t$, the
ground state of the half-filled electron lattice is a Kondo
insulator, with a singlet formed at each site 
between each localized moment (blue) and conduction electron (red), see
Eq.~(\ref{KI}).  (b) When electrons are removed from the
Kondo-insulator, 
the resulting positively charged 
holes carry the spin of the underlying local moment
and can hop from site to site, forming a heavy fermion metal.
(c) When two holes are created
above the diagonal sites A and B of the underlying Shastry--Sutherland lattice,
they form a singlet with binding energy $-J_1$.}

(ii) Now consider the large $K$ limit where $J_{K}\gg J_{1}\gg t$.  If we have
exactly one electron per site, the state formed is a Kondo insulator,
with Kondo singlets at each site (Fig. \ref{fig.strong} (a)):
\beq\label{KI}
|\text{KI}\rangle =   \prod_i \frac{1}{\sqrt{2}} \left(  c\dg_{i\up}
f\dg_{i\dn}- c\dg_{i\dn} f\dg_{i\up} \right) |0\rangle
\eeq
By removing electrons from this
state, one creates the ``holes'' (Fig. \ref{fig.strong}(b)) that form the excitations of the large
Fermi surface state, $\tilde{h}\dg_{j\s} \equiv -c_{j-\s}\sgn ( {\sigma })$, so that
\beq\label{hole}
\tilde{h}\dg_{_j\s} |\text{KI}\rangle = \frac{1}{\sqrt{2}}  
f\dg_{j\,\sigma} \prod_{i\neq j} 
\frac{1}{\sqrt{2}} 
\left(  c\dg_{i\up} f\dg_{i\dn} - c\dg_{i\dn} f\dg_{i\up}
\right)|0\rangle.
\eeq
 Creating a hole in the half-filled lattice (i.e. an empty
site) destroys the Kondo singlet and thus has an energy cost $J_K$ which can be
absorbed into the chemical potential, $\tilde{\mu} = \mu - J_K$. The holes can
hop from site to site, as illustrated in Fig.~\ref{fig.strong}(b).
 While generically, there are no spin dimers present, if two holes happen to
come together across the diagonal of a plaquette, then a dimer can form on the
underlying Shastry--Sutherland $f$-spin lattice, as shown schematically in
Fig.~\ref{fig.strong}(x). The energy of such configuration will be lowered by an
amount of order $-J_{1}$ in the first order of the perturbation theory. 
In this way, the effective Hamiltonian at large $K$ will take the form
\begin{equation}
H_{eff}' = \sum_{\kk\s} \tilde{\eps}_k \hks\dg \hks +
J_1\sum_{\boxslash,\boxbslash} \boldsymbol{\tilde{\sigma}}_k\cdot\boldsymbol{\tilde{\sigma}}_l
\end{equation}
where $\hks\dg $ creates a heavy $f$-hole, as per Eq.~(\ref{hole}), with
dispersion
$\tilde{\epsilon}_{\bk }= + t (\cos k_{x}+\cos k_{y})-\tilde{\mu}$ and spin
$\boldsymbol{\tilde{\sigma}}_i = \frac{1}{2}(\tilde{h}_{i\alpha}\dg
\boldsigma_{\alpha\beta}\tilde{h}_{i\beta})$.
Because of the effective antiferromagnetic coupling on the alternating
diagonals, this ``large Fermi surface'' state will also be susceptible to the
$d_{xy}$
superconducting instability.

\begin{figure}[!tb]
\begin{center}
{\includegraphics[width=0.7\textwidth]{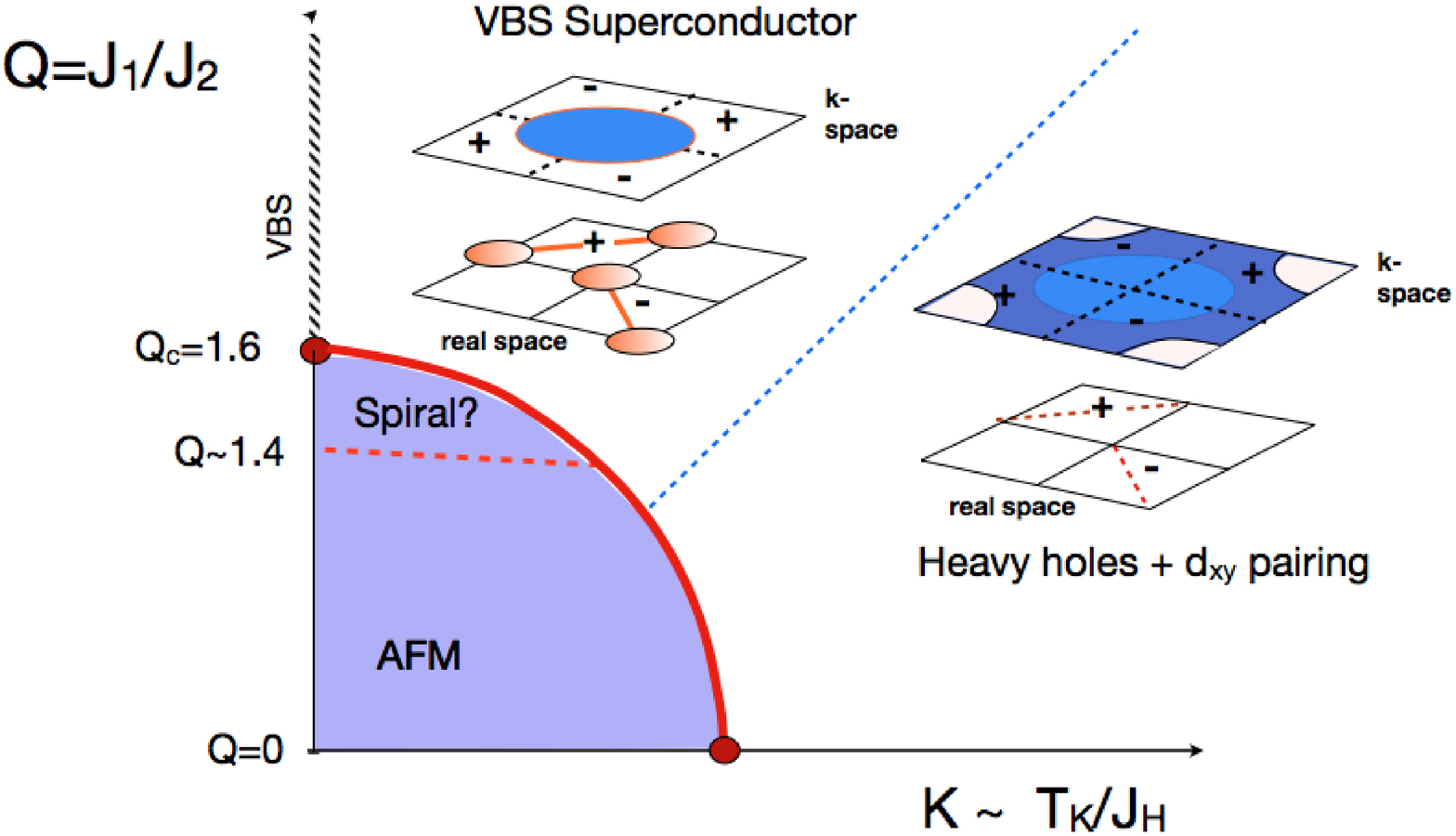}}
\end{center}
\caption{(Colour online) Schematic phase diagram of the
Shastry--Sutherland--Kondo model. The
frustration axis denotes $Q=J_1/J_2$. At half-filling, there is a metal
insulator transition from the insulating dimer phase on the left into a
$d$-wave superconductor with a large Fermi surface on the right. At a generic
filling, both phases are metallic and the transition becomes a crossover.
 } 
\label{Fig.SSK}
\end{figure}

In this way, strong coupling arguments suggest the prevalence of a
$d_{xy}$ superconductor at both large and small values of $K$.
Adiabaticity can then be used to argue that unless anything unforeseen
occurs, the two superconducting states are connected across the phase diagram,
shown schematically in Fig.~\ref{Fig.SSK}.

A similar set of arguments can be advanced using a large $N$ treatment 
of the Shastry--Sutherland Kondo lattice.
Let us again consider the limit of infinite frustration $J_2=0$ ($Q\to\infty$)
when the dimer phase is stable.  
The dimer phase can be captured by using the fermionic representation of spins, 
$\mathbf{S}_i = \frac{1}{2} f\dg_{i\alpha}\boldsigma_{\alpha\beta}f_{i\beta}$,
subject to the occupancy constraint $n_f=1$. We introduce an anomalous operator 
\beq
B\dg_{ij}=\sum_\sigma \sgn(\sigma) f\dg_{i\sigma}f\dg_{j -\sigma}
\eeq
which creates a singlet pair of $f$-electrons on the diagonal links of the
Shastry--Sutherland model. While physical spins are described by SU(2)
group, the above expression can be easily generalized to the spins that belong
to symplectic Sp($N$) group, with $\sigma=-\frac{N}{2},\dots \frac{N}{2}$.
In the large-$N$ mean field theory, the dimer phase is
described by the non-zero expectation value $\Delta =J_1 \langle
B_{ij} 
\rangle/N$, and
the spin Hamiltonian can be written as
\beq
\tilde{H}_{SS} = -\sum_i (\Delta_\boxslash B\dg_{ij} + h.c.) - \sum_k (\Delta _\boxbslash
B\dg_{kl} + \text{h.c.}) + N \frac{|\Delta _\boxslash|^2}{J_1} + 
N \frac{|\Delta _\boxbslash|^2}{J_1},
\eeq
where by symmetry one expects the dimer averages to be the same on the
alternate diagonals, up to an arbitrary phase: $\Delta _\boxbslash = \Delta _\boxslash
e^{i\phi}$.

The heavy Fermi liquid ground state, on the other hand, can be
characterized by non-zero on-site expectation
value of hybridization $V_i=-J_K \sum_\sigma \langle c\dg_{i\sigma}
f_{i\sigma} \rangle/N$. The full large-$N$ mean-field Hamiltonian then becomes:
\beq
H^{MF} = \sum_{\kk \s} \eps_k \cks\dg \cks + \tilde{H}_{SS}
+ \sum_{i\s} (V f\dg_{i\s} c_{i\s} + \text{h.c.})  + N\frac{|V|^2}{J_K}.
\eeq
In the limit $K\ll 1$, the dimer phase dominates, with non-zero expectation
value of $\Delta _\boxslash$ and $\Delta _\boxbslash$. In the opposite limit of strong
Kondo interaction, $V$ will acquire a non-zero expectation value. 
Generically, both order parameters may co-exist in a certain region of the
phase diagram. 

It is easy to show that such coexistence describes a superconducting
instability of the conduction electrons within this large-$N$ mean field
theory. This is related
to the fact that the product of the two mean-field order parameters $V_i V_j
B_{ij}$ translates into the average:
\beq
V_i V_j \Delta _{ij} \propto 
 \langle (c\dg_{i}f_{i})\cdot \hat B\dg_{ij}\cdot (c\dg _{j} f_{j})\rangle 
\propto  \langle c\dg _{i\s}c\dg _{j-\s}{\rm sgn} (\sigma ) \rangle,
\eeq
i.e. it describes pairing of conduction electrons on the diagonals of the
Shastry--Sutherland lattice.
It is clear from the above strong coupling arguments that the extended $s$-wave
pairing will be at disadvantage compares with the $d$-wave pairing, and so the
above anomalous average will describe $d_{xy}$ pairing.
 
The resulting phase diagram of the Shastry--Sutherland Kondo lattice model is
shown schematically in Fig.~\ref{Fig.SSK}.
Current experimental realization of the Shastry--Sutherland
Kondo lattice, including Yb$_2$Pt$_2$Pb compound~\cite{kim08} and its
cerium analogs~\cite{kim2010}, lie either in the antiferromagnetic
region, or on the edge of the paramagnetic/spin liquid region of the
phase diagram.  The possibility of a superconducting ground-state,
co-existing with dimer order is an fascinating possibility for future work.  
More studies of the Shastry--Sutherland Kondo model are needed, and its full
understanding, through theory and experiment, appears to be an important
step towards future understanding of the interplay of frustration and Kondo
effects in heavy fermion systems.


\section{Experimental considerations}\label{experiment}

\fg{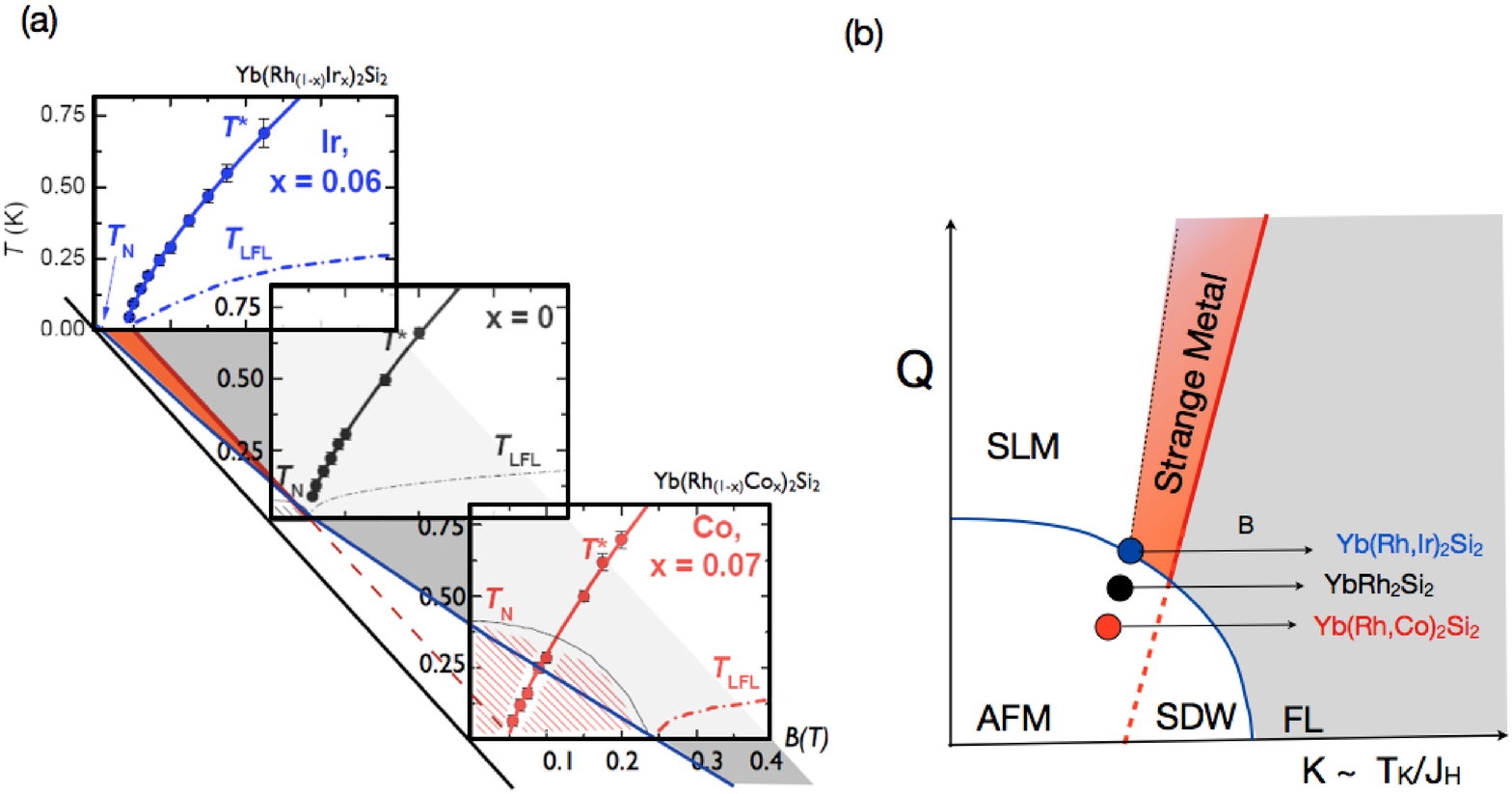}{Fig.exp}{(a) Temperature-field phase diagrams for Co- and Ir-doped
\yrs after Ref.~\onlinecite{sven2009}, superposed onto the $QK$ diagram. 
(b) Interpretation of the experiments~\cite{sven2009} on
doped \yrs within the $QK$ diagram, the arrows showing the effect of applying
magnetic field. }



There is mounting experimental  evidence for the importance of magnetic
frustration in heavy fermion physics. While the
examples of $f$-electron metals on strongly geometrically frustrated lattices
are very rare, such as e.g. 
the pyrochlore compound \priro, there are indications that
a number of
compounds crystallizing in an \emph{a priori} non-frustrated tetragonal lattice,
become \emph{de facto} frustrated thanks to competing RKKY interactions between
the nearest and next nearest neighbours. The existence of a spiral
antiferromagnetic phase~\cite{park-nature06} in CeRhIn$_5$ is a particularly
striking
example of such a competition, on a
par with apparent absence of any long-range magnetic
order in the sister superconducting compound CeCoIn$_5$ (although the N\'eel
order can be induced by doping with Cd~\cite{niklas07}). In another well-studied heavy fermion material, \yrs, the
very tiny N\'eel temperature $T_N\approxeq70$~mK\cite{trovarelli00},
which can be further
suppressed by doping with Ir~\cite{sven2009}, also points to the importance
of magnetic frustrations in this system.
 
It is in these tetragonal lattice systems that the small- to large- Fermi
surface transition has been most clearly established.
There is strong experimental evidence from the Hall effect
measurements that applying magnetic field to pure~\cite{paschen04} and
doped~\cite{sven2009} \yrs leads to a jump in the Fermi surface
volume. A similar phenomenon occurs when hydrostatic pressure is
applied to CeRhIn$_5$, when a clear change of the Fermi surface
topology is seen in the de Haas--van Alphen measurements in the
field-induced normal state~\cite{shishido}. 
As we argued above, our $QK$ phase diagram provides reason to believe that
the small-to-large Fermi surface transition should be present also in the
spin-liquid phase, independent of the antiferromagnetic phase boundary. This
point-of-view is also supported by other theories~\cite{senthil04, rech06}. 

  Indeed, recent measurements on doped~\cite{sven2009} and
pressurized~\cite{yrs-pressure} \yrs unambiguously show that the small-to-large
Fermi surface transition is separate from the conventional antiferromagnetic
quantum critical point. This is illustrated in Fig.~\ref{Fig.exp}, in which we
attempted to put the existing data on Co- and Ir-doped \yrs onto the
theoretical $QK$ diagram. It has to be kept in mind that generically, changing
experimental parameters, such as pressure, doping or magnetic field,
may involve simultaneous changes along both the Doniach and
frustration axes in the $QK$ phase diagram. 

In the case of doped \yrs, the data
suggest~\cite{sven2009} that applying magnetic field drives the system along the
horizontal $K$-axis, as shown with arrows in  
schematic phase diagram in Fig.~\ref{Fig.exp}b). Crucially, the Co-doped
material exhibits first the jump in the Hall coefficient at smaller field, and
then an AFM to paramagnet transition, which suggests that it is likely to be
deeper inside the AFM phase then pure \yrs, as shown by the red circle in
Fig.~\ref{Fig.exp}(b).
By contrast, the Ir-doped material first undergoes a transition into a
paramagnet, and then develops a large Fermi surface characteristic of a heavy
Fermi liquid. The strange phase that lies in between cannot thus be a
conventional Fermi liquid, and on the schematic phase diagram
Fig.~\ref{Fig.exp}(b) we denote it as a ``strange metal'' phase, forming a wedge
along the line of small- to large- Fermi surface transition. It is tempting
therefore to place the Ir-doped \yrs closer to the spin-liquid phase (blue
circle in Fig.~\ref{Fig.exp}b) then the pure compound (black circle). 

Intriguingly, the ``strange metal'' phase on the $QK$ phase diagram may 
actually be realized in the recently discovered \ybal
compound~\cite{ybal,ybal-scaling}. This material appears to be quantum
critical without any external tuning, as characterized by anomalous critical
exponents and $T/B$ scaling. Unless nature has fortuitously placed \ybal
right at a quantum critical point, this 
material is most probably an example of  ``strange metal'' phase,
shown as the coloured wedge in Fig.~\ref{Fig.exp}(b). 

Furthermore, the ``quantum critical phase'' in \ybal appears to be
unstable to application of magnetic field.  A tiny field, 
comparable to the Earth's magnetic field, is sufficient to drive the
material back into a Fermi liquid.
That magnetic field is a relevant perturbation to this  critical phase is a very
important observation which deserves further experimental and theoretical work.
It is worth noting that structurally very similar compound, $\alpha$-YbAlB$_4$,
shares the same features as the $\beta$-phase in
its thermodynamic properties (susceptibility, specific
heat)~\cite{ybal-scaling} above $T^*\sim 2$~K, yet has a
Fermi liquid ground state. The possibility that by applying hydrostatic pressure
one may be able to drive \ybal into a Fermi liquid phase, just as
$\alpha$-YbAlB$_4$ is at ambient pressure, is very intriguing and deserved
experimental attention.

One of the conclusions arising from theoretical considerations presented
earlier, is that the spin liquid state (or valence bond solid, as in the
Shastry--Sutherland case) may be reached
either by increasing the Doniach ratio $K=T_K/J^H$ or by increasing the amount of
frustration $Q$. In fact it has
been recently proposed~\cite{custers10} that a number of experimental
systems, such as YbAgGe and Ge-doped \yrs may lie in the vicinity of
the point where the three phases cross in the $QK$ phase
diagram in Fig.~\ref{Fig.exp}(b).
In this case it should be possible,  in principle, to drive
these systems into a spin liquid state by applying either (chemical) pressure or
magnetic field.

\section{Conclusions and outlook}

Our paper has discussed and illustrated
how a unified consideration of the
effects of spin zero-point fluctuations (denoted by ``Q'') and the Kondo
screening (encoded by ``K'') leads to a two dimensional
global phase diagram of the Kondo lattice. In particular, general
arguments based on the $QK$ diagram lead
us to conclude that the antiferromagnetic phase
boundary and the $f$-delocalization transition may be partially, if not totally
independent of one another. This possibility was first observed theoretically
a number of years ago 
\cite{senthil04}, and if correct, may prove rather liberating  for the
theoretical community, allowing us to split the problem of quantum
criticality into two separate studies of $f$-electron magnetization and
localization. 

Previously, the various non-Fermi liquid properties observed in the
vicinity of the AFM phase transition of heavy fermion compounds, such
as the quasi-linear resistivity $\rho \sim \rho_0 + T^{\alpha }$ ($\alpha \sim
1$) and the logarithmic temperature dependence of the specific heat
($C_{V}/T\sim - \ln T)$ were phenomenologically associated with a
single transition. Viewed from the new perspective, 
 it is tempting to associate
many of these features with $f$-electron delocalization,
rather than the development of magnetism, \emph{per se}. 
Senthil\cite{senthilcrit} has proposed that the small to large Fermi surface
transition will involve a critical Fermi surface, reminiscent of that found
in Luttinger liquids.  It would be fascinating indeed, if such a scenario
separated the spin and heavy fermi liquids.

To provide a concrete example, we have considered above the
Shastry--Sutherland Kondo (SSK) lattice model, which is believed
to be experimentally realized~\cite{kim08} in the heavy fermion compound Yb$_2$Pt$_2$Pb.
Its beauty lies in the fact that while being an example of a fully frustrated
spin model (provided $Q\gg 1$), the ground state of the Shastry--Sutherland
model is known exactly. This lifts the uncertainty as regards the ground state,
which is otherwise present in virtually all examples of geometrically frustrated
spin models. In particular, the SSK model lends itself to
large-$N$ treatment. The combination of the latter, together with
strong-coupling arguments, allowed us to sketch qualitatively the $QK$ phase
diagram of the model.
Ironically, our theory predicts that, with the exception of the special case of
half-filling, the quantum critical point is most likely replaced by the $d$-wave
superconducting phase which prevails over the entire zero-temperature phase
diagram. Nevertheless, this model provides a useful insight into the 
interplay of frustration and Kondo
effect in heavy fermion systems and warrants further detailed studies, both
theoretical and experimental.

Historically, the advancement of  a theory  of classical
criticality benefited from a whole host of developments -- the solution of
the  2D Ising model, the development of classical Monte Carlo methods
that opened the physics to numerical analysis, 
the idea of the renormalization group, the
abstraction of phase transitions in terms of a continuum 
Ginzburg--Landau theory, and lastly the development of new kinds of controlled
mathematical treatments, such as the $\epsilon$ and the $1/N$ expansion.

It would seem likely that a parallel set of developments are required
in order for us to understand the nature of quantum criticality in
heavy fermion materials.  A number of useful directions
seem to present themselves:
\begin{itemize}

\item Exploration of fully frustrated  Kondo lattices. Kondo
generalizations of the  Shastry-Sutherland and kagom\'e lattice
may offer the
hope of characterizing the $f$-electron
delocalization transition without the complications of magnetism. Present
fermionic theories~\cite{paul06, pepin07, pepin08} of the Kondo lattice cannot
describe the transition into the magnetically ordered state.

\item  The development of the idea of ``local quantum
criticality''\cite{qmsi,qmsi2} to incorporate soft charge 
degrees of freedom as a possible approach to $f$-electron
delocalization. To date, ideas of local quantum criticality have
focussed on the notion that the soft critical excitations are the
magnetic spins themselves. Yet, when spins of 
local magnetic moments delocalize, 
the critical spin excitations
must also involve critical \emph{charge} degrees of freedom.
The inclusion of the latter,
possibly as zero energy charged fermions, into a locally quantum
critical theory may prove fruitful in this respect. 

\item  One of the unsolved problems, is to find a controlled expansion
that unifies magnetism with the Kondo effect. The large-$N$ Schwinger
boson approach\cite{rech06} appears to offer hope in this
respect. The Schwinger boson large-$N$ limit of the Kondo lattice
involves self-consistent integral equations which have been solved for
the one and two impurity Kondo models, but which remain unsolved for the
Kondo lattice. The solutions of these integral equations at the
$f$-delocalization transition could shed considerable light on the
nature of heavy fermion quantum criticality. 

\item The development of radically new kinds of large $N$ expansion based on an
idea from String theory sometimes called ``holography'' and the 
``Anti-de Sitter / Conformal Field Theory'' (AdS/CFT) correspondence.
\end{itemize}

We should like to end with a few comments on this last approach.
The AdS/CFT correspondence 
is a conjecture from String theory\cite{maldacena}, 
which maps the large-$N$ limit of certain conformal 
field theories in $d$ dimensions
onto a  {\sl classical theory}  of waves 
moving in the gravitational field of a black hole in a higher-dimensional curved
space (Anti-de Sitter space). The conjecture can be
written in a deceptively over-simplified form as 
\begin{eqnarray}
\left\langle 
{\rm T}\exp\left[\int \ud^d x\, {\phi_0 (x)}{\hat \psi}(x)\right]\right\rangle
_{\color{red}QFT} &=& e^{-S_{\color{red}grav}{[\phi_0}]}\cr\cr
{\rm lim}_{r\rightarrow \infty}\phi_E(r;x)&\rightarrow &\phi_0(x)
\end{eqnarray}
where $\phi_{0} (x)$ is a source term that couples to the physical
fields in the $d$-dimensional field theory, while in the
($d+1$)-dimensional gravity theory, $\phi_{E} (r,x)$ is a classical
field whose asymptotic ($r\rightarrow \infty $) behaviour converges to
the value $\phi_{0} (x)$.

This approach holds the mathematically tantalizing promise that
by solving radial  {\sl one-particle}
wave equations in the higher-dimensional  world  (albeit rather
complex radial equations, for particles in the curved space-time of a 
charged black hole), one can extract the 
universal, i.e. quantum critical, physics of 
interacting fermions in our lower-dimensional world.
The method has already shown its utility in
deriving properties of certain model non-Fermi liquids with a critical
Fermi surface\cite{liu09,zaanen,mcgreevy2009}. Nevertheless, there are
there are a number of important hurdles to be overcome 
before this approach proves 
its usefulness to heavy fermion physics, in
particular:
\begin{itemize}
\renewcommand{\labelitemi}{$\diamond$}
\item  In the current theories, the fermions 
in the ``strange metal''  produced at the boundary of the Anti-de Sitter
space are spinless - with a purely orbital coupling to magnetic fields
and consequently, a purely {\sl diamagnetic
magnetic} ($\chi<0$) susceptibility. 
Can the Zeeman/spin physics, with a proven Pauli paramagnetism ($\chi >0$) 
be impl<emented in future implementations of the AdS/CFT
correspondence?( One idea, suggested in \cite{siads} is to introduce
an additional global symmetry. )

\item There is currently no Fermi liquid solution to the AdS/CFT gravity
equations. If the quantum critical behaviour seen in condensed matter
physics is universal, we expect that the 
cross-over from a critical metal to a Fermi liquid is
part of that universal theory. This should motivate future  attempts
to drive a cross-over to Fermi liquid behaviour in the AdS/CFT approach.

\end{itemize}
\renewcommand{\labelitemi}{$\bullet$}

We should like to acknowledge fruitful  discussions connected with
this work with Vic Alexandrov,
Meigan Aronson, Jeroen Custers, Rebecca Flint, Hilbert von Lohneysen, Silke
Paschen, Hong Liu, Frank Steglich, Matt Strassler, Scott Thomas and Qimiao Si. 
This work was supported by NSF grant DMR 0907179. 


\bibliographystyle{apsrev}

\end{document}